\documentclass[12pt]{article}
\usepackage{amsthm,amsfonts,amsmath,amssymb,amscd,mathrsfs,url,graphicx}

\title{Multi-Time Wave Functions}
\author{
Matthias Lienert\footnote{Department of Mathematics, Rutgers University,
	110 Frelinghuysen Road, Piscataway, NJ 08854-8019, USA. E-mail: m.lienert@rutgers.edu},\ \ 
S\"oren Petrat\footnote{Department of Physics, Jadwin Hall, Princeton University, 
	Washington Road, Princeton, NJ 08544-0708, USA. E-mail: spetrat@princeton.edu},\ \ and 
Roderich Tumulka\footnote{Department of Mathematics, Rutgers University,
	110 Frelinghuysen Road, Piscataway, NJ 08854-8019, USA. E-mail: tumulka@math.rutgers.edu}
}
\date{February 15, 2017}

\begin{document}
\maketitle

\newcommand{\be}{\begin{equation}}
\newcommand{\ee}{\end{equation}}
\newcommand{\Hilbert}{\mathscr{H}}
\newcommand{\conf}{\mathcal{Q}}
\newcommand{\Fock}{\mathscr{F}}
\renewcommand{\Re}{\mathrm{Re}}
\renewcommand{\Im}{\mathrm{Im}}
\newcommand{\RRR}{\mathbb{R}}
\newcommand{\R}{\mathbb{R}}
\newcommand{\CCC}{\mathbb{C}}
\newcommand{\C}{\mathbb{C}}
\newcommand{\scp}[2]{\langle #1|#2 \rangle}
\newcommand{\vq}{\boldsymbol{q}}
\newcommand{\vy}{\boldsymbol{y}}
\newcommand{\vx}{\boldsymbol{x}}
\newcommand{\diag}{\mathrm{diag}}
\newcommand{\sM}{\mathscr{M}}
\newcommand{\sS}{\mathscr{S}}
\newcommand{\CST}{\mathscr{C}}

\begin{abstract}
In non-relativistic quantum mechanics of $N$ particles in three spatial dimensions, the wave function $\psi(\vq_1,\ldots,\vq_N,t)$ is a function of $3N$ position coordinates and one time coordinate. It is an obvious idea that in a relativistic setting, such functions should be replaced by $\phi((t_1,\vq_1),\ldots,(t_N,\vq_N))$, a function of $N$ space-time points called a multi-time wave function because it involves $N$ time variables. Its evolution is determined by $N$ Schr\"odinger equations, one for each time variable; to ensure that simultaneous solutions to these $N$ equations exist, the $N$ Hamiltonians need to satisfy a consistency condition. This condition is automatically satisfied for non-interacting particles, but it is not obvious how to set up consistent multi-time equations with interaction. For example, interaction potentials (such as the Coulomb potential) make the equations inconsistent, except in very special cases. However, there have been recent successes in setting up consistent multi-time equations involving interaction, in two ways: either involving zero-range ($\delta$ potential) interaction or involving particle creation and annihilation. The latter equations provide a multi-time formulation of a quantum field theory. The wave function in these equations is a multi-time Fock function, i.e., a family of functions consisting of, for every $n=0,1,2,\ldots$, an $n$-particle wave function with $n$ time variables. These wave functions are related to the Tomonaga--Schwinger approach and to quantum field operators, but, as we point out, they have several advantages.
\end{abstract}

\section{Introduction}
\label{sec:intro}

\begin{figure}[h]
\begin{center}
\includegraphics[width=0.7 \textwidth]{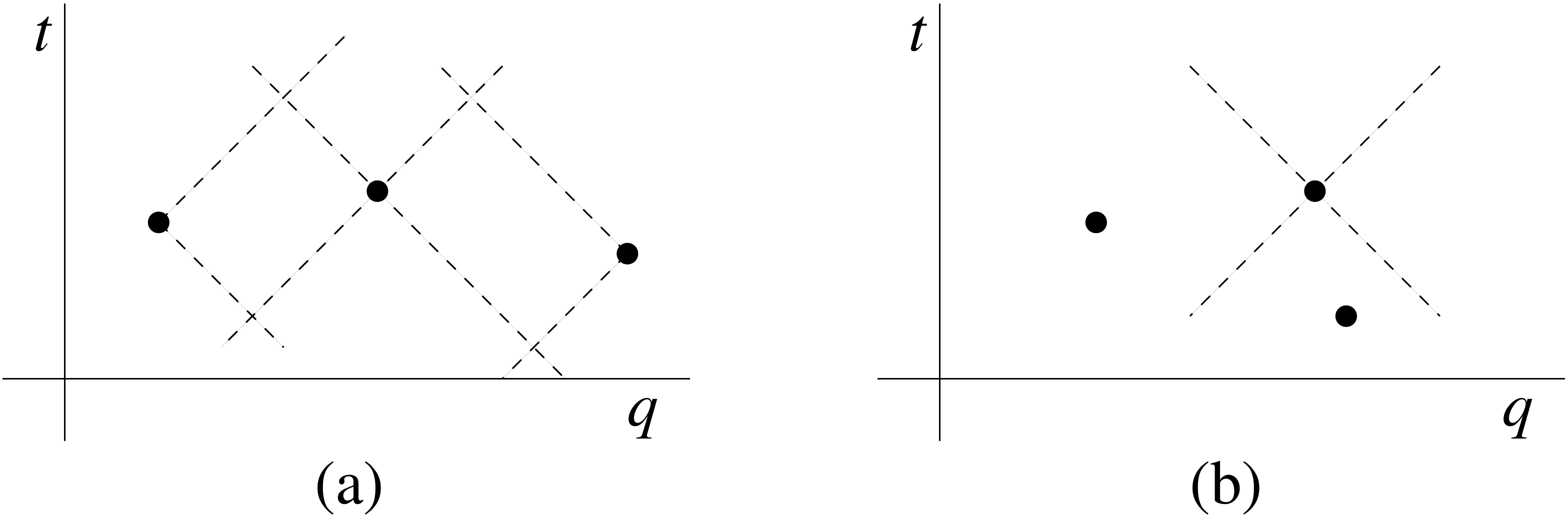}
\end{center}
\caption{Example of (a) a spacelike configuration of 3 points (shown as bullets), (b) a non-spacelike configuration. Both examples are shown in Minkowski space-time with light cones (dashed) drawn at $45^\circ$.}
\label{fig:spacelike}
\end{figure}

Multi-time wave functions arise naturally when considering a particle-position representation of a quantum state in a relativistic setting. They were first introduced by Dirac in 1932 \cite{dirac:1932} and studied to some extent in the 1930s \cite{dfp:1932,bloch:1934}, but not comprehensively. The basic idea is that, in a relativistic space-time, ordinary $N$-particle wave functions
\be\label{psi}
\psi\bigl(\vq_1,\vq_2,\ldots,\vq_N,t\bigr)
\ee
with $\vq_j\in\RRR^3$
require the choice of a reference frame because they refer to the positions of several particles \emph{at the same time} $t$. An alternative that does not require a choice of reference frame is to consider a wave function
\be\label{phi}
\phi\Bigl( (t_1,\vq_1), (t_2,\vq_2), \ldots, (t_N,\vq_N) \Bigr)
\ee
that is a function of $N$ space-time points $x_j=(t_j,\vq_j)$ and thus of $N$ time variables, called a \emph{multi-time wave function}. We call an $N$-tuple of space-time points a \emph{space-time configuration}, or simply a \emph{configuration}. The function $\phi$ is a covariant object: It does not require the choice of any coordinate system on space-time $\sM$ if we regard it as a function $\phi: \sM^N \to S$, with $S$ a suitable spin space (or $S=\CCC$ in the spinless case, or $S$ a bundle of spin spaces if $\sM$ is curved). More precisely, $\phi$ will often be defined only on the \emph{spacelike configurations}, that is, on the set $\sS_N$ of those $N$-tuples $(x_1,\ldots,x_N)$ of space-time points $x_j\in\sM$ for which any two are spacelike separated or equal; see Figure~\ref{fig:spacelike}. Note that $\sS_N$ is also defined in a covariant way, and is also $4N$-dimensional.

The relation between $\psi$ and $\phi$ is simple: In the reference frame to which $\psi$ refers, set all time variables in $\phi$ equal to obtain $\psi$,
\be\label{psiphi}
\psi\Bigl(\vq_1,\vq_2,\ldots,\vq_N,t\Bigr) = \phi \Bigl( (t,\vq_1), (t,\vq_2), \ldots, (t,\vq_N) \Bigr)\,.
\ee 
Put differently, this means that $\psi$ is the restriction of $\phi$ to the \emph{simultaneous configurations} relative to the chosen reference frame.

As we will explain in more detail below, $\phi$ is usually also directly related to detection probabilities according to the \emph{curved Born rule}: If we place detectors along a spacelike hypersurface $\Sigma$, then the probability distribution on $\Sigma^N$ of the detected configuration has density (relative to the volume defined by the 3-metric on $\Sigma$) given by
\be\label{curvedBorn}
\rho(x_1,\ldots,x_N) = \bigl|\phi(x_1,\ldots,x_N)\bigr|^2
\ee
for any $x_1,\ldots,x_N \in \Sigma$ and with $|\cdot |^2 = |\cdot|^2_\Sigma$ understood appropriately: for example, for Dirac wave functions with spin space $S=(\CCC^4)^{\otimes N}$,
\be\label{squaredef}
|\phi|^2 = \overline\phi \: \Bigl[\gamma^{\mu_1}n_{\mu_1}(x_1) \otimes \cdots \otimes \gamma^{\mu_N} n_{\mu_N}(x_N)\Bigr] \: \phi~,
\ee
where $n_\mu(x)$ is the future unit normal vector to $\Sigma$ at $x\in\Sigma$. In words, the inner product in spin space (and thus the norm) depends on the Lorentz frame, and we need to use the local frame tangent to $\Sigma$.

A time evolution law for $\phi$ is a law that determines $\phi$ on its entire domain from initial data. The appropriate initial datum in a given reference frame specifies the values of $\phi$ at those configurations for which all $t_j=0$ while the $\vq_j$ are arbitrary; in other words, the initial datum is $\psi(t=0)$. The kind of evolution analogous to the Schr\"odinger equation (we set $\hbar=1$)
\be\label{Schr}
i \frac{\partial \psi}{\partial t} = H\psi
\ee
is a system of PDEs comprising one equation per time variable,
\be\label{multi}
i \frac{\partial \phi}{\partial t_j} = H_j \phi\,,
\ee
called \emph{multi-time (Schr\"odinger) equations}. The multi-time equations we are considering are linear equations, so that linear combinations of solutions are solutions. The chain rule and \eqref{psiphi} then imply the single-time Schr\"odinger equation \eqref{Schr} with
\be
H= \sum_j H_j
\ee
at space-time configurations with $t_1=t_2=\ldots = t_N$. A central issue about multi-time equations that does not arise for the ordinary Schr\"odinger equation is that the $H_j$ need to fulfill a \emph{consistency condition}, or else the equations \eqref{multi} cannot be simultaneously satisfied, or can only for special initial conditions. Therefore, whenever we propose a system of multi-time equations, we need to prove their consistency. For non-interacting particles, the consistency condition is automatically fulfilled, and in fact a unique solution $\phi$ exists, not only on the spacelike configurations, but on all of $\sM^N$ \cite{schweber:1961}. In contrast, to set up consistent multi-time equations with interaction is challenging. Apart from special examples \cite{DV82b,DV85,CVA:1983,CVA:1997} (and early successes with fields \cite{bloch:1934,tomonaga:1946,schwinger:1948}, more below), this was successfully done only recently \cite{pt:2013c,pt:2013d,lienert:2015a,LN:2015}; we will elucidate below how. In all of these examples, the multi-time equations are remarkably simple, see Eqs. \eqref{eq:twotime}, \eqref{bdyconds}, and \eqref{part_Ham} below.

In quantum field theory (QFT), the single-time wave function $\psi$ can often be taken to be an element of Fock space, and thus a function on $\conf=\bigcup_{N=0}^\infty (\RRR^3)^N$ (with the union understood as a disjoint union), the configuration space of a variable number of particles. The corresponding multi-time wave function $\phi$ is defined on a subset of $\bigcup_{N=0}^\infty \sM^N$ (with $\sM$ the space-time, say $\sM=\RRR^4$), viz., the set of spacelike configurations $\sS=\bigcup_{N=0}^\infty \sS_N$. The multi-time equations are then an infinite system of coupled partial differential equations of the type \eqref{multi}, with interaction implemented via creation and annihilation terms in the $H_j$. Since the $N$-particle sector $\phi^{(N)}$ of $\phi$ (i.e., the part of $\phi$ on $\sS_N$) has $N$ time variables, there are $N$ equations for it; the creation and annihilation terms involve $\phi^{(N+1)}$ and $\phi^{(N-1)}$. In Section~\ref{sec:qft} we provide an explicit example of such a set of equations which has been shown to be consistent. There is also a simple connection of the multi-time wave function to an expression involving the field operators in the Heisenberg picture, and to the Tomonaga-Schwinger approach. In fact, under suitable conditions, all these three approaches can be translated into each other. 

While our motivation comes from the wish for a manifestly covariant particle-position representation of the quantum state, we mention that Elze \cite{El16a,El16b,El17} has recently used multi-time wave functions for a different purpose in connection with certain discrete action principles called Hamiltonian cellular automata: Elze found that for an $N$-particle system with $N>1$, such action principles for a multi-time wave function can yield a physically more reasonable time evolution (after setting all times equal) than for a single-time wave function.

Another application of multi-time wave functions finds particular use in multi-time equations \emph{without interaction}: it concerns detection probabilities on a \emph{timelike} hypersurface $\tilde\Sigma$, corresponding to detectors waiting for the particles to arrive. Since different particles can arrive at the detectors at different times, the joint distribution $\tilde\rho$ of the space-time points of detection naturally involves several time variables. While in the case with interaction, the computation of $\tilde\rho$ involves collapses of the wave function for any (attempted) detection, $\tilde\rho$ can be computed more easily in the case without interaction, in fact directly from the multi-time wave function $\phi$, also at non-spacelike $(x_1\ldots x_N)$, according to
\be
\tilde\rho(x_1\ldots x_N) =
\overline{\phi(x_1\ldots x_N)}\: \bigl[\gamma^{\mu_1}\tilde{n}_{\mu_1}(x_1)\otimes \cdots \otimes \gamma^{\mu_N}\tilde n_{\mu_N}(x_N)\bigr] \: \phi(x_1\ldots x_N)
\ee
with $\tilde n_\mu(x)$ the outward unit normal vector to $\tilde\Sigma$ at $x$, at least in the following two cases: (i)~for ideal hard detectors modeled by an absorbing boundary condition on $\tilde\Sigma$ \cite{detect-dirac}; and (ii)~in the scattering regime \cite{DT04}, where detectors are placed along a very distant surface in space and stay there, so that the particles coming out of the scattering process do not interact because of the great distance.

The remainder of this article is organized as follows. In Section~\ref{sec:unitarity}, we explain how the multi-time approach is related to a Hilbert space framework, and how the multi-time wave function relates to detection probabilities. In Section~\ref{sec:consistency}, we elucidate the need for and form of consistency conditions. In Section~\ref{sec:potentials}, we summarize results showing that interaction potentials make multi-time equations inconsistent. In Section~\ref{sec:zerorange}, we describe a consistent model with zero-range interaction, and in Section~\ref{sec:qft} consistent models in QFT.

\section{Hilbert spaces, unitarity, and detection probabilities}\label{sec:unitarity}

\subsection{Hilbert spaces and unitarity}\label{sec:Hilbert}

Unitarity plays a crucial role in the structure of quantum physics. Given a multi-time wave function $\phi$ on its natural domain $\mathscr{S}$, one cannot, however, insert $N$ arbitrary time variables and expect that the integral of $|\phi|^2$ over the space variables yields unity. The reason is that $\phi(t_1,\vq_1,\ldots,t_N,\vq_N)$ is not defined for all configurations, but only for spacelike configurations. Instead, the integral of $|\phi|^2$ over a spacelike Cauchy hypersurface $\Sigma$ yields unity. More precisely, let us define $\phi_\Sigma$ from $\phi$ through the appropriate ``restriction to $\Sigma$,'' i.e., by considering only configurations on $\Sigma$:
\be
\phi_\Sigma(q) := \phi(q),~~~q \in \Sigma^N.
\label{phisigma}
\ee
Here $N$ can be either fixed, in the case of a fixed number of particles, or take several values referring to different sectors of Fock space for a variable particle number. Now the integral of $|\phi_\Sigma|^2$ over $\conf_\Sigma=\Sigma^N$ (or $\conf_\Sigma=\bigcup_{N=0}^\infty \Sigma^N$) equals 1, as it must for the curved Born rule \eqref{curvedBorn} to make sense. Thus, $\phi_\Sigma$ lies in the appropriate Hilbert space $\Hilbert_\Sigma$ associated with $\Sigma$, e.g., for $N$ Dirac particles,
\be
\Hilbert_\Sigma = \Hilbert_{N,\Sigma} =S_{\pm} L^2\bigl(\Sigma^N,(\CCC^4)^{\otimes N}\bigr)
\ee
with $S_{\pm}$ the (anti-)symmetrizer assuming the particles are bosons (fermions);\footnote{According to the spin-statistics connection, Dirac particles must be fermions, but for toy models we may equally well consider bosonic symmetry.} the inner product is
\be
\scp{f}{g} = \int_{\Sigma^N} d^3x_1\cdots d^3x_N\, \overline{f(x_1\ldots x_N)}\, \Bigl[ \gamma^{\mu_1} n_{\mu_1}(x_1) \otimes \cdots \otimes \gamma^{\mu_N} n_{\mu_N}(x_N)\Bigr] \, g(x_1\ldots x_N)\,,
\ee
so that $\|f\|^2=\scp{f}{f}$ equals the integral of the probability density \eqref{squaredef}.

Since we can take $\phi_\Sigma$ as an initial datum, have the multi-time equations determine $\phi$ on all spacelike configurations, and then consider $\phi_{\Sigma'}$ on any other spacelike Cauchy hypersurface $\Sigma'$, we obtain a time evolution operator
\be
\label{usigma}
U_\Sigma^{\Sigma'}: \Hilbert_\Sigma \rightarrow \Hilbert_{\Sigma'},~~~\phi_\Sigma \mapsto \phi_{\Sigma'}
\ee
that is unitary for the multi-time equations considered here. These unitaries satisfy the composition laws $U_{\Sigma'}^{\Sigma''} U_{\Sigma}^{\Sigma'}=U_\Sigma^{\Sigma''}$ and $U_\Sigma^\Sigma = I$. Furthermore, they provide the translation between the multi-time wave function $\phi$ and the Tomonaga-Schwinger equation, as we will discuss in Section~\ref{sec:qft}.

This family of unitaries is largely equivalent to the multi-time evolution. More precisely, if a family $(\phi_\Sigma)_\Sigma$ consisting of one element in each $\Hilbert_\Sigma$ is given, these functions fit together as a single function $\phi$ on the set $\sS$ of spacelike configurations according to \eqref{phisigma} if and only if
\be
\label{objectivitycond}
\phi_\Sigma(q) = \phi_{\Sigma'}(q)
\quad \text{whenever }q \in \Sigma^N \cap (\Sigma')^N\,.
\ee
This relation is satisfied for $U_{\Sigma}^{\Sigma'}$ obtained from the Tomonaga-Schwinger equation in relevant examples. 

In the case of a variable number of particles, the multi-time wave function becomes a function on $\bigcup_{N=0}^\infty \sS_N \subset \bigcup_{N=0}^\infty \sM^N$ called a ``multi-time Fock function'' \cite{pt:2013c}. It can be represented as a sequence of $N$-particle multi-time wave functions $\phi^{(N)}$,
\be
\phi = \bigl( \phi^{(0)}, \phi^{(1)}, \phi^{(2)}, \ldots \bigr),
\ee
where $\phi^{(0)} \in \CCC$. We write $\phi(q) = \phi^{(N)}(q)$ if $q = (x_1,\ldots,x_N)$. $\phi_\Sigma$ then is an element of the Fock space
\be
\Hilbert_\Sigma = \bigoplus_{N=0}^\infty \Hilbert_{N,\Sigma} 
\ee
with
\be
\bigl\| \phi_\Sigma \bigr\|^2 = \sum_{N=0}^\infty \bigl\| \phi^{(N)}_\Sigma \bigr\|^2 = 1\,.
\ee

\subsection{Detection probabilities and the curved Born rule}\label{sec:Born}

A full proof of the curved Born rule is the subject of work in progress \cite{LPT:2017}; here we briefly outline what needs to be proved, as well as prior results.

\begin{figure}[h]
\begin{center}
\includegraphics[width=0.7 \textwidth]{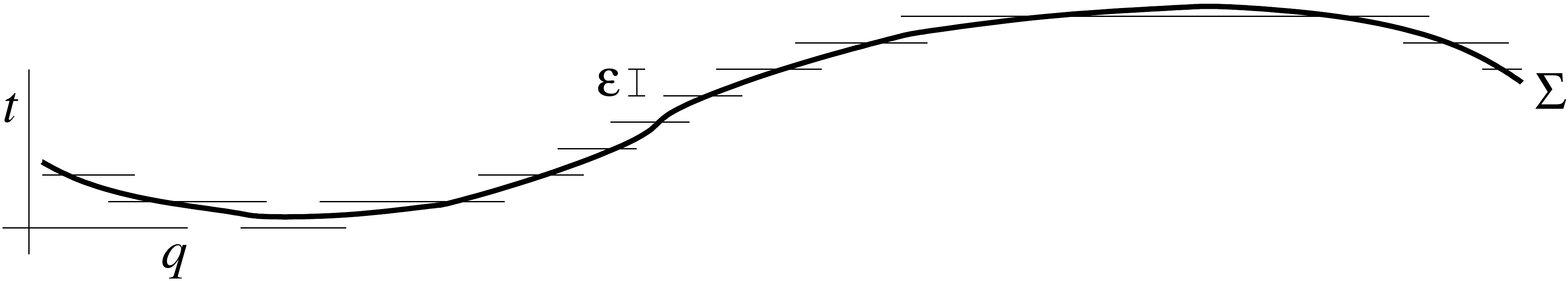}
\end{center}
\caption{Construction for computing the detection probability distribution on a curved spacelike hypersurface $\Sigma$: $\Sigma$ gets approximated by horizontal pieces with temporal distance $\varepsilon$. The length of the pieces is chosen such that every timelike curve intersects at least one piece.}
\label{fig:discrete}
\end{figure}

The unitarity of $U_{\Sigma}^{\Sigma'}$ entails that $|\phi_\Sigma|^2$ integrates up to 1 and thus qualifies as a probability distribution on $\conf_\Sigma=\Sigma^N$ or $\conf_\Sigma=\bigcup_{N=0}^\infty \Sigma^N$---it is the natural candidate for a curved Born rule. However, this rule cannot simply be postulated, because the usual Born rule in any one fixed Lorentz frame, together with the appropriate collapse rule, already determines the joint probability distribution of the detection events for detectors that we place at different times, including detectors that we place along any $\Sigma$. Specifically, if we approximate $\Sigma$ in the given Lorentz frame by horizontal pieces of hypersurfaces as in Figure~\ref{fig:discrete} with temporal discretization $\varepsilon$, then the usual Born and collapse rules apply to the horizontal pieces as a kind of iterated position measurements with repeated collapse, one after every attempted detection. 

In the limit $\varepsilon\to 0$ we obtain a distribution on $\conf_\Sigma$, and the claim is that this distribution coincides with $|\phi_\Sigma|^2$. Here it is relevant that wave functions do not propagate faster than light, and that interaction terms in the Hamiltonian do not provide faster-than-light interaction.

\begin{figure}[h]
\begin{center}
\includegraphics[width=0.9 \textwidth]{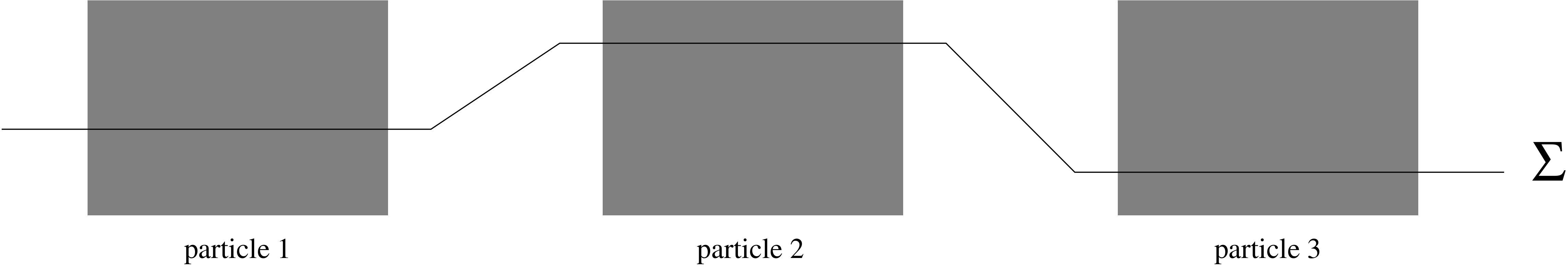}
\end{center}
\caption{Bloch's \cite{bloch:1934} result concerns particles confined to spacelike separated regions (gray) and hypersurfaces that are horizontal in each region.}
\label{fig:Bloch}
\end{figure}

A preliminary result in this direction was already obtained by Bloch \cite{bloch:1934} (see also \cite[Sec. 2.3]{lienert:2015c} for a discussion in English): He derived the curved Born rule in the case that $N$ particles are confined to spacelike separated regions as in Figure~\ref{fig:Bloch} (so they cannot interact), and $\Sigma$ is horizontal within each region.

\section{Consistency of evolution equations}
\label{sec:consistency}

Multi-time evolution equations are not necessarily consistent. One needs to ensure that the many simultaneous equations \eqref{multi} do not contradict each other. Consider first the case in which the number $N$ of particles (and thus of time variables) is fixed, and the $H_j$ are time-independent self-adjoint operators on a Hilbert space $\Hilbert$. Regarding $\phi$ as a function $\RRR^N \rightarrow \Hilbert,~(t_1,\ldots,t_N) \mapsto \phi(t_1,\ldots,t_N)$, and for given initial data
\be \label{initialdata}
\phi\bigl(t_1 = 0,\ldots,t_N= 0\bigr) = \phi_0\,,
\ee
the order of first time-evolving $\phi_0$ in $t_j$ and then in $t_k$ or the other way around must be irrelevant, i.e., the following diagram has to commute:
\be
\CD
\phi(0, 0) @>e^{-iH_j t_j}>> \phi(t_j, 0)\\
@Ve^{-iH_k t_k}VV @VVe^{-iH_k t_k} V\\
\phi(0,t_k) @>e^{-iH_j t_j}>> \phi(t_j,t_k).
\endCD
\ee
In words, $e^{-iH_j t_j}$ and $e^{-iH_k t_k}$ have to commute for all $t_j$ and $t_k$, which happens if and only if the $H_j$ commute (in the spectral sense) \cite[thm. VIII.13]{reedsimon1}:
\be \label{timelessconsistency}
[H_j, H_k] = 0~~~ \forall j,k.
\ee
In that case, 
\be
\phi(t_1,\ldots,t_N) = e^{-iH_1 t_1} \cdots e^{-i H_N t_N} \phi(0,\ldots,0)\,.
\ee
If the $H_j$ depend on time, \eqref{timelessconsistency} has to be replaced by the following consistency condition \cite{pt:2013a}:
\be \label{consistency}
\Bigl[i \partial_{t_j} - H_j, i \partial_{t_k} - H_k\Bigr] = 0~ \forall j,k.
\ee
This condition has been shown to be both necessary and sufficient in the case of bounded operators $H_j$ on $\Hilbert$ \cite{pt:2013a} and some other cases \cite{pt:2013c}. To find a rigorous proof of necessity and sufficiency in general remains a task for future work. We conjecture that \eqref{consistency} is the appropriate consistency condition also when the operators $H_j$ are general differential expressions, such that the multi-time equations remain first-order partial differential equations in the times $t_j$. This includes the cases of multi-time equations \eqref{multi} (a) which are defined on a sub-domain of $\RRR^{4N}$, and (b) with a variable number of time coordinates. Case (a) occurs, e.g., as multi-time wave functions are naturally only defined on $\mathscr{S}$ (see the examples \cite{pt:2013c,lienert:2015a}). Case (b) is the typical situation in quantum field theory when formulated in the particle-position representation \cite{pt:2013c}. Then, the expression for $H_j$ is no longer an operator on Hilbert space; it still is a differential expression, as we will discuss below. 

Furthermore, we conjecture that a given single-time dynamics \eqref{Schr} with finite propagation speed and local interactions can always be extended to yield a unique (consistent) multi-time evolution. This is supported by the examples \cite[proof of thm. 8]{pt:2013a} and \cite[Sec. 5.4]{pt:2013c} and shall be the subject of future work.

\section{Inconsistency of interaction potentials}
\label{sec:potentials}

The consistency condition \eqref{consistency} is quite restrictive. Two of us obtained a no-go theorem about interaction potentials in \cite{pt:2013a} which was further extended in \cite{ND:2016}. Here, ``interaction potentials'' are understood as arbitrary smooth matrix-valued functions $V_j(x_1,\ldots,x_N) : \R^{4N} \rightarrow (\CCC^4)^{\otimes N}$ in 
\be
H_j = H_j^0 + V_j(x_1,\ldots,x_N)\,,
\ee
where
\be\label{free_Dirac}
H_j^0 = \sum_{k=1}^3 i \gamma^0_j\gamma_j^k \frac{\partial}{\partial x_j^k} + m \gamma_j^0
\ee
is the free Dirac Hamiltonian (we set $c=1$) acting on the coordinates and spin indices of the $j$-th particle, and $\gamma_j^\mu$ denotes the Dirac gamma matrix $\gamma^\mu$ acting on the $j$-th tensor factor of $(\CCC^4)^{\otimes N}$. (Note that the superscript 0 in $\gamma^0$ means the timelike component of $\gamma^\mu$, whereas in $H^0$ it means something else: the free Hamiltonian.)

The combined results of \cite{pt:2013a,ND:2016} then state that the only Poincar\'{e} invariant potentials $V_j$ which satisfy the consistency condition \eqref{consistency} are $V_j \equiv 0$. Similar theorems for the cases $H_j^0 = - \Delta_j$ and $H_j^0$ given by arbitrary first-order differential operators can also be obtained \cite{pt:2013a}. If we drop the requirement about Poincar\'{e} invariance, then potentials satisfying \eqref{consistency} can be found, but these seem artificial and consequently only of mathematical interest \cite{ND:2016}. Note that the proof in \cite{pt:2013a} was carried out for smooth potentials only, but we expect the result to hold for singular potentials as well, e.g., the Coulomb potential $|\vq|^{-1}$.

This result raises the question: How can interaction be achieved in multi-time equations other than via potentials? Two answers, based on zero-range interaction and on particle creation/annihilation, will be provided in Sections \ref{sec:zerorange} and \ref{sec:qft}. Other approaches have been suggested in \cite{DV82b,DV85,CVA:1983,CVA:1997} (see also \cite{lienert:2015b}); another notable approach is based on integral equations for multi-time wave functions \cite[appendix A]{lienert:2015c}; in fact, the well-known Bethe-Salpeter equation \cite{bethe_salpeter} belongs to this class.

\section{Relativistic zero-range interactions}
\label{sec:zerorange}

While the above no-go theorem excludes interaction potentials that are functions, it does not exclude $\delta$ potentials, also known as zero-range interactions. It is known \cite{AGHKH88} from non-relativistic quantum mechanics that zero-range interactions can be implemented rigorously by means of a boundary condition on the wave function at those configurations for which two particles meet. This clearly avoids the use of interaction potentials.

We now describe an example \cite{lienert:2015a,lienert:2015c} of a consistent multi-time evolution with zero-range interaction for two massless Dirac particles in 1+1 dimensions, $\sM=\RRR^2$. The reasons for setting up the model in this way are the following. In the relativistic case, we need to choose a relativistic Hamiltonian such as the Dirac Hamiltonian. For the latter, it is known \cite{Sve81}, however, that zero-range interactions exist in 1+1 but not in higher dimensions. In the massless case, the dynamics becomes particularly simple and even explicitly solvable such that new mathematical techniques not relying on the time-less functional analytic Hilbert space picture become available. These are needed for a manifestly covariant treatment of the model.

The multi-time wave function in this case is a map
\be
\phi : \mathscr{S} \subset \RRR^2 \times \RRR^2 \rightarrow \CCC^2 \otimes \CCC^2 \simeq \CCC^4,~~~ (t_1,z_1,t_2,z_2) \mapsto \phi(t_1,z_1,t_2,z_2)
\ee
and the multi-time equations are given by the free 1+1-dimensional Dirac equations with a boundary condition. The free equations read, with the notation $x_j=(t_j,z_j)$,
\be
\label{multitimedirac}
i \gamma_j^{\mu} \partial_{j, \mu} \, \phi (x_1,x_2) ~=~ 0 , ~~j= 1, 2
\ee
in covariant notation, or
\begin{eqnarray}
i \frac{\partial}{\partial t_1} \phi(t_1, z_1, t_2, z_2) &=& - i \, \sigma_3 \otimes 1_2 \frac{\partial}{\partial z_1} \, \phi(t_1, z_1, t_2, z_2), \nonumber\\
i \frac{\partial}{\partial t_2} \phi(t_1, z_1, t_2, z_2) &=& - i \, 1_2 \otimes \sigma_3 \frac{\partial}{\partial z_2} \,\phi(t_1, z_1, t_2, z_2).
\label{eq:twotime}
\end{eqnarray}
in Hamiltonian form. Here, $1_2$ stands for the $2\times 2$ unit matrix,
\be
\gamma^0 = \sigma_1 = \left( \begin{array}{cc} 
0 & 1 \\ 1 & 0
\end{array} \right), ~~~ \gamma^1 = \sigma_1 \sigma_3 = \left( \begin{array}{cc} 
0 & -1 \\ 1 & 0
\end{array} \right),
\ee
and $\sigma_i,~i = 1,2,3$ denote the Pauli matrices. The boundary conditions are prescribed (as limits) on the set of collision configurations,
\be
\mathscr{C} :=\{ (t_1,z_1,t_2,z_2) \in \R^2 \times \R^2 : t_1 = t_2,~z_1=z_2 \}
\label{coincidencepoints},
\ee
and one particular example of a suitable boundary condition is given by (denoting the spin components of $\phi$ as $\phi_i,~i=1,2,3,4$)
\begin{align}
\phi_2(t,z-0,t,z+0) = e^{-i \theta} \phi_3(t,z-0,t,z+0),\nonumber\\
\phi_2(t,z+0,t,z-0) = e^{+i \theta} \phi_3(t,z+0,t,z-0),
\label{bdyconds}
\end{align}
where $\theta \in (-\pi, \pi]$ is a phase.

Initial data are given at equal times as in \eqref{initialdata}; they have to satisfy the boundary condition as well. The main results are: the multi-time evolution is consistent and can be defined in a rigorous way, the model is interacting in the sense that a generic initial product wave function $\phi_0$ becomes entangled with time, both multi-time equations as well as boundary conditions are Lorentz invariant, and the model is compatible with anti-symmetry for indistinguishable particles. Heuristically, the boundary conditions \eqref{bdyconds} correspond to a spin-dependent $\delta$-potential $(\pi - \theta) \diag(0,1,-1,0) \delta(z_1-z_2)$ at equal times \cite{LN:2015}.
Finally, the Dirac tensor current
\be
j^{\mu \nu}_\phi (x_1,x_2) = \overline{\phi}(x_1,x_2) \gamma^\mu \otimes \gamma^\nu \phi(x_1,x_2)
\ee
is conserved, 
\be
\partial_{x_1^\mu} j^{\mu\nu}(x_1,x_2) = \partial_{x_2^\nu} j^{\mu\nu}(x_1,x_2) =0\,,
\ee
which, together with the boundary conditions \eqref{bdyconds}, ensures the unitarity of $U_{\Sigma}^{\Sigma'}$.

The model has also been extended to $N$ particles in 1+1 dimensions in \cite{LN:2015}, and there are strong indications that non-zero masses will not change the results. An extension to higher dimensions, however, does not seem feasible as then the dimension of $\mathscr{C}$ is too low for boundary conditions to have impact on the dynamics. In conclusion, the results show that interacting dynamics for multi-time wave functions in one spatial dimension can be achieved in a rigorous and manifestly Lorentz invariant way.

\section{Quantum field theory}
\label{sec:qft}

Another way of implementing interaction in the multi-time framework is by particle creation and annihilation, i.e., by considering models from quantum field theory. We report here mainly about the results of \cite{pt:2013c,pt:2013d}. Interaction by particle creation naturally suggests itself when we are looking for relativistic theories and expect that interaction should not take place faster than light; besides, perhaps surprisingly, also the consistency condition of multi-time equations (relativistic or not) pushes us, since it excludes interaction through potentials, to considering particle creation. 
The formulation of models from QFT in terms of multi-time wave functions can be regarded as a new representation of QFT, a multi-time Schr\"odinger picture particle position representation. As such, it provides an alternative approach to fully relativistic formulations of QFT such as the Tomonaga-Schwinger formalism and quantum fields in the Heisenberg picture. We will later argue that these three pictures are in fact equivalent, in the sense that each can be translated into the others under suitable conditions. (Some of the statements in this direction we show in very general terms, while others we show for specific examples, but we conjecture to hold also for more general models.) 

Nevertheless, an advantage of the multi-time framework is that as a mathematical object multi-time wave functions are simpler, since they are locally just functions of finitely many variables, and their evolution is determined by a coupled system of PDEs. Furthermore, it is possible to introduce a cut-off in the multi-time framework \cite{p:2010,pt:2013a}, but not in the Tomonaga-Schwinger picture. This is so because there is no analogue of spacelike hypersurfaces with a cut-off, but there are versions of the set of all almost spacelike configurations that take into account a finite range of the interaction. On the other hand, the multi-time approach as we present it here has the limitation that it is tied to a particle-position representation of the involved particles (using Fock space), and cannot be applied to a field representation (in which $\psi$ is a functional of a function on 3-space). In fact, Dirac, Fock, and Podolsky \cite{dfp:1932} considered a model of quantum electrodynamics in a particle representation for the electrons and a field representation for the radiation. As a multi-time formulation, they proposed one time variable for each electron and one time variable for the field. That led them to considering the field on a horizontal hyperplane, although it was a main motivation for multi-time wave functions to avoid being tied to configurations on horizontal hyperplanes; that is why we do not follow their suggestion here.

In the examples we consider here, we take all particles to be Dirac particles; photons could be included by taking a photon wave function to be a complexified Maxwell field \cite{p:2010}. We leave aside issues of the right choice of position observable and make no attempt to remove or redefine the negative energy solutions. We also leave aside the ultraviolet divergence problem and calculate in a non-rigorous way.

\subsection{Multi-time equations with particle creation and annihilation}

We describe in detail a simple toy model, the ``emission-absorption model'' \cite{pt:2013c}. 
It involves fermionic $x$-particles that can emit and absorb bosonic $y$-particles, so that the $x$-particle number is conserved and the $y$-particle number is not. We first define the single-time version of the model. Its Hilbert space is given by the tensor product of two Fock spaces,
\begin{equation}
\Hilbert = \Hilbert_x \otimes \Hilbert_y, ~~~~~\text{with}~~ \Hilbert_{x / y} = \bigoplus_{N=0}^\infty S_{x / y} L^2(\RRR^3,\CCC^4)^{\otimes N},
\end{equation}
where $S_x$ is the antisymmetrization operator, and $S_y$ the symmetrization operator.\footnote{This choice is again contrary to the spin-statistics connection, since we take both $x$- and $y$-particles to be spin-$\frac{1}{2}$ Dirac particles. The spin-statistics \emph{theorem} does not apply here, since our Hamiltonian \eqref{H_onetime} is unbounded from below.} In other words, the wave function $\psi$ in the $(M,N)$-particle sector takes values in the spinor space $(\CCC^4)^{\otimes M} \otimes (\CCC^4)^{\otimes N}$, i.e., it could be explicitly written as $\psi_{r_1\ldots r_M,s_1\ldots s_N}(x^{3M}, y^{3N})$, abbreviating $(x^{3M},y^{3N}) = (\vx_1,\ldots,\vx_M,\vy_1,\ldots,\vy_N)$ with $\vx_j,\vy_k \in \RRR^3$ for all $j,k$. However, in a given expression, we usually only indicate the indices that other operators than the identity act on. The time evolution of the wave function $\psi_t \in \Hilbert$ is given by the Schr\"odinger equation
\begin{equation}\label{Schr_onetime}
i \partial_t \psi_t = H\psi_t, ~~~~~\text{with}~~ H = H_x + H_y + H_{\mathrm{int}}.
\end{equation}
In order to write down the interaction, we introduce the usual (spinor valued) creation and annihilation operators $a^\dagger, a$ for the $x$-particles and $b^\dagger, b$ for the $y$-particles. These satisfy the (anti-)commutation relations
\begin{equation}
\{ a_r(\vx), a^\dagger_{r'}(\vx') \} = \delta_{rr'} \delta(\vx-\vx')~~~\text{and}~~ [ b_s(\vy), b^\dagger_{s'}(\vy') ] = \delta_{ss'} \delta(\vy-\vy'),
\end{equation}
and all other combinations are zero ($a$ and $b$ operators commute). Explicitly written out in position space, they read
\begin{align}
\bigl(a_r(\vx)\,\psi\bigr)(x^{3M},y^{3N}) 
&= \sqrt{M+1}\; (-1)^M\, \psi_{r_{M+1}=r}\bigl((x^{3M},\vx),y^{3N}\bigr)\label{adef}\\
\bigl(a_r^\dagger(\vx)\,\psi\bigr)(x^{3M},y^{3N}) 
&= \frac{1}{\sqrt{M}} \sum_{j=1}^M (-1)^{j+1} \, \delta_{rr_j}\,\delta^3(\vx_j-\vx)\, \psi_{\widehat{r_j}}\bigl(x^{3M}\setminus \vx_j,y^{3N}\bigr)\,,\\
\bigl(b_s(\vx)\,\psi\bigr)(x^{3M},y^{3N}) 
&= \sqrt{N+1}\; \psi_{s_{N+1}=s}\bigl(x^{3M},(y^{3N},\vx)\bigr)\label{bdef}\\
\bigl(b_s^\dagger(\vx)\,\psi\bigr)(x^{3M},y^{3N}) 
&= \frac{1}{\sqrt{N}} \sum_{k=1}^N \delta_{ss_k}\, \delta^3(\vy_k-\vx)\,\psi_{\widehat{s_k}}\bigl(x^{3M},y^{3N}\setminus \vy_k\bigr)\, ,
\end{align}
where $\hat{\cdot}$ means that the variable is omitted and $(x^{3M}\setminus \vx_j) := (\vx_1,\ldots,\vx_{j-1},\vx_{j+1},\ldots,\vx_M)$. Then the contributions to the Hamiltonian are
\begin{align}\label{Hams}
H_x &= \int d^3\vx\, a^\dagger(\vx)H^0a(\vx), ~~ H_y = \int d^3\vx\, b^\dagger(\vx)H^0b(\vx), \nonumber\\
H_{\mathrm{int}} &= \int d^3\vx\, a^{\dagger}(\vx)\Big( g^*\cdot b(\vx) + g\cdot b^\dagger(\vx) \Big) a(\vx),
\end{align}
where $H^0$ is the free Dirac Hamiltonian as in \eqref{free_Dirac}, $g \in \CCC^4$ is a fixed spinor, and the summation over the spinor indices is implicitly understood in the above expressions ($\cdot$ denotes the inner product in spinor space). In order to compare better with the multi-time equation that we are going to introduce next, we write down the Hamiltonian in the position representation:
\begin{align}\label{H_onetime}
\big( H \psi \big)(x^{3M},y^{3N}) &= \sum_{j=1}^M H^0_{x_j} \psi(x^{3M},y^{3N}) + \sum_{k=1}^N H^0_{y_k} \psi(x^{3M},y^{3N}) \nonumber\\
&\quad + \sqrt{N+1} \sum_{j=1}^M \sum_{s_{N+1}=1}^4 g^*_{s_{N+1}} \psi_{s_{N+1}}\big( x^{3M}, (y^{3N},\vx_j)\big) \nonumber\\
&\quad + \frac{1}{\sqrt{N}} \sum_{j=1}^M\sum_{k=1}^N g_{s_k} \delta(\vy_k-\vx_j) \psi_{\widehat{s_k}}(x^{3M},y^{3N} \setminus \vy_k).
\end{align}

In this setting, the multi-time wave function $\phi$ is a spinor-valued function on the set of all spacelike configurations $\sS=\bigcup_{M,N=0}^\infty \sS_{M,N}$, where $\sS_{M,N}$ is the set of all spacelike configurations of $M$ $x$- and $N$ $y$-particles. Similar to before we write $(x^{4M},y^{4N}) = (x_1,\ldots,x_M,y_1,\ldots,y_N)$ with $x_j,y_k \in \RRR^4$ for all $j,k$. Then the multi-time evolution equations are given by
\begin{align}\label{mt_emab}
i \frac{\partial \phi}{\partial x_j^0} = H_{x_j}\phi, ~~~~ i \frac{\partial \phi}{\partial y_k^0} = H_{y_k}\phi,
\end{align}
with
\begin{align}\label{part_Ham}
H_{x_j}\phi(x^{4M},y^{4N}) &= H^0_{x_j} \phi(x^{4M},y^{4N}) + \sqrt{N+1} \sum_{s_{N+1}=1}^4 g^*_{s_{N+1}} \phi_{s_{N+1}}\big( x^{4M}, (y^{4N},x_j)\big) \nonumber\\
&\quad + \frac{1}{\sqrt{N}} \sum_{k=1}^N G_{s_k}(y_k-x_j) \phi_{\widehat{s_k}}(x^{4M},y^{4N} \setminus y_k), \nonumber\\[4mm]
H_{y_k}\phi(x^{4M},y^{4N}) &= H^0_{y_k} \phi(x^{4M},y^{4N}),
\end{align}
where $G:\RRR^4\to\CCC^4$ is a Green's function, i.e., the solution to
\begin{equation}\label{Green_multi}
i \frac{\partial G}{\partial y^0} = H_{y}^0 G ~~~~~\text{with}~~ G(0,\vy) = g \delta(\vy)
\end{equation}
for the fixed spinor $g \in \CCC^4$. One could also rewrite the multi-time equations in a covariant notation by multiplying by $\gamma^0$ and bringing the free Hamiltonian to the left-hand side; then they take the form
\be
i\gamma_{x_j}^\mu \partial_{x_j^\mu} \phi + m_y \phi = \ldots\quad \text{and} \quad
i\gamma_{y_k}^\mu \partial_{y_k^\mu} \phi + m_y \phi = 0\,.
\ee 
Note that for equal times $x_j^0=y_k^0=t$ for all $j,k$, the equations \eqref{mt_emab} indeed reduce to the single-time Schr\"odinger equation \eqref{Schr_onetime} with Hamiltonian \eqref{H_onetime}. In fact, the multi-time equations are little more than the terms in the one-time $H$ \eqref{H_onetime} grouped into terms associated with each particle---remarkably simple. Furthermore, the solution $\phi$ to \eqref{mt_emab} has the same permutation symmetry as the initial datum, but now as a permutation of \emph{space-time} points:
\begin{align}
\phi_{r_{\sigma(1)}\ldots r_{\sigma(M)},s_1\ldots s_N}\big(x_{\sigma(1)}, \ldots, x_{\sigma(M)},y^{4N}\big) &= (-1)^{\sigma} \phi_{r_1\ldots r_M,s_1\ldots s_N}\big(x_1, \ldots, x_M,y^{4N}\big), \nonumber\\
\phi_{r_1\ldots r_M,s_{\sigma(1)}\ldots s_{\sigma(N)}}\big(x^{4M},y_{\sigma(1)}, \ldots, y_{\sigma(N)}\big) &= \phi_{r_1\ldots r_M,s_1\ldots s_N}\big(x^{4M},y_1, \ldots, y_N\big)
\end{align}
for all permutations $\sigma$, with $(-1)^{\sigma}$ the sign of the permutation.

It was shown in \cite{pt:2013c} that also in this case with a variable number of time variables, the commutator condition \eqref{consistency} is necessary and sufficient for the consistency of the multi-time equations. It turns out that for the operators from \eqref{part_Ham} these commutators indeed vanish on all spacelike configurations (including collision configurations where some $x_j$ or $y_k$ are equal). Thus, on a non-rigorous level, the equations \eqref{mt_emab} possess a unique solution on $\sS$, given initial conditions for equal times. In the two cases when $m_y=0$ or $g^*\gamma^0 g = 0$, the commutators vanish on \emph{all} configurations (also non-spacelike ones), but we believe these to be exceptional cases due to the simplicity of the model.

A few remarks about the multi-time model seem in order.
\begin{itemize}
\item When setting up multi-time equations starting from a single-time model with Hamiltonians such as \eqref{H_onetime}, there is usually some freedom in how to distribute the different terms among the different multi-time equations. In \eqref{H_onetime}, since the creation term contains a sum over all $x$- and $y$-particles, one can attribute its summands either to $H_{x_j}$ or to $H_{y_k}$. That is, one could also set up the multi-time equations \eqref{mt_emab} with 
\begin{align}\label{part_Ham_alt}
H_{x_j}\phi(x^{4M},y^{4N}) &= H^0_{x_j} \phi(x^{4M},y^{4N}) + \sqrt{N+1} \sum_{s_{N+1}=1}^4 g^*_{s_{N+1}} \phi_{s_{N+1}}\big( x^{4M}, (y^{4N},x_j)\big), \nonumber\\
H_{y_k}\phi(x^{4M},y^{4N}) &= H^0_{y_k} \phi(x^{4M},y^{4N}) + \frac{1}{\sqrt{N}} \sum_{j=1}^M G_{s_k}(y_k-x_j) \phi_{\widehat{s_k}}(x^{4M},y^{4N} \setminus y_k).
\end{align}
However, it turns out that the multi-time equations with \eqref{part_Ham_alt} are equivalent to the ones with \eqref{part_Ham} on the set $\sS$ of spacelike configurations. This fact is rather surprising: How can two sets of equations that give different, non-equivalent expressions for certain partial derivatives of $\phi$ (viz., for $\partial \phi/\partial x_j^0$) be equivalent? This has to do with the set $\sS$ of spacelike configurations: If $\phi$ were defined on the set $\bigcup_{M,N=0}^\infty \sM^{M+N}$ of \emph{all} configurations (spacelike or not), then the equations \eqref{part_Ham_alt} would \emph{not} be equivalent to \eqref{part_Ham} because one could simply compute $\partial \phi/\partial x_j^0$ and see whether it agrees with the right-hand side of the first equation in \eqref{part_Ham} or that in \eqref{part_Ham_alt}. However, for $\phi$ defined on $\sS$ there are certain configurations where $\partial \phi/\partial x_j^0$ cannot be computed: the configurations where an $x$-particle and a $y$-particle meet, $x_j=y_k$. There, varying $x_j^0$ while keeping $y_k^0$ fixed would lead out of $\sS$, so $\partial \phi/\partial x_j^0$ is not defined, whereas $(\partial/\partial x_j^0 + \partial/\partial y_k^0) \phi$ is. And the crucial term $G_{s_k}(y_k-x_j)$ vanishes in $\sS_N$ \emph{except at precisely those configurations} where $x_j=y_k$. At those configurations, the PDEs \eqref{mt_emab} are understood as determining the derivatives of $\phi$ that \emph{are} defined, and that is why different choices of $H_{x_j}$ and $H_{y_k}$ can determine these derivatives in the same way, and thus define the same time evolution of $\phi$.

\item The model would also be consistent on $\sS$ if the wave function had been chosen symmetric under exchange of $x$-particles, i.e., if the $x$-particles were bosonic. However, it is interesting to note that the model would \emph{not} be consistent on $\sS$ if the wave function was antisymmetric in the $y$-particles, i.e., if the $y$-particles were fermionic. (It would then not be consistent in the special cases $m_y=0$ or $g^*\gamma^0 g = 0$ either; in fact, the commutators \eqref{consistency} for $x_i\neq x_j$ would be non-zero at \emph{all} configurations.) In other words, the model is only consistent if the fermion number is conserved (which is believed to be one of the fundamental conservation laws of the Standard Model).

\item The operator $H_{x_j}$ from \eqref{part_Ham} or \eqref{part_Ham_alt} is \emph{not} an operator on a Hilbert space, because it involves changing a time variable, viz., setting $y_{N+1}^0 = x_j^0$. It is a perfectly fine operator acting on $\phi$, but cannot be understood as an operator on a Hilbert space. That is not surprising keeping in mind (i)~that $H_{x_j}\phi$ provides $i\partial \phi/\partial x_j^0$ for a specific sector $\phi^{(M,N)}$ but depends on neighboring sectors $\phi^{(M,N\pm 1)}$; and (ii)~that the functions $\phi$ with fixed time coordinates and arbitrary space coordinates do not form a Hilbert space, as discussed in Section~\ref{sec:Hilbert} above. The multi-time equations do define evolution operators $U_{\Sigma}^{\Sigma'}$ between Hilbert spaces $\Hilbert_\Sigma$ and $\Hilbert_{\Sigma'}$, as elucidated in Section~\ref{sec:TS} below.

\item Initial data that determine $\phi$ can, in fact, be specified on any spacelike hypersurface $\Sigma$ by specifying $\phi$ on $\conf_\Sigma = \bigcup_{M,N=0}^\infty \Sigma^{M+N}$, i.e., for all configurations on $\Sigma$.

\item The multi-time equations \eqref{mt_emab} are not fully Lorentz invariant because they involve the choice of a fixed spinor $g$ and the only Lorentz invariant $4$-spinor is the zero spinor (which would make the model interaction-free). However, if the $y$-particles had integer spin, $g$ could be replaced by a Lorentz-invariant object, and the equations would be fully covariant. 

\item As mentioned before, the equations \eqref{mt_emab} are actually not rigorously defined since they contain $\delta$ distributions in the creation terms via the Green's function $G$ (which is a distribution). This is an instance of the ultraviolet divergence in QFT. The equations \eqref{mt_emab} could be defined in a rigorous sense by introducing an ultraviolet cut-off, i.e., replacing the $\delta$ distribution in \eqref{Green_multi} with some localized function $\varphi$. This is described in more detail in \cite{p:2010}, where the evolution equations with such a cut-off are explicitly written down, and a proof for the consistency and existence and uniqueness of solutions on a (not Lorentz invariant) subset of the set of spacelike configurations is sketched. However, this obviously breaks the Lorentz invariance of the equations. Since the multi-time formalism is about finding a fundamentally relativistic invariant quantum theory, we chose to proceed with formal calculations. (However, note that for those equations where a renormalization scheme works, one could also apply this to the multi-time equations.) It will be of interest to explore whether and how multi-time equations can be set up for Hamiltonians for which creation and annihilation terms are defined by means of boundary conditions instead of $\delta$ functions \cite{ibc2}.
\end{itemize}

\subsection{Relation to field operators in the Heisenberg picture}

There is a simple relation between the multi-time wave function $\phi$, i.e., the solution to \eqref{mt_emab}, and an expression involving creation and annihilation operators in the Heisenberg picture. In the latter, the state vector $\Psi$ is fixed and only the operators are subject to the dynamics, i.e., we define for $x=(t,\vq)$ that
\begin{equation}
a(x) = e^{iHt}a(\vq)e^{-iHt}, ~~~~~~~ b(x) = e^{iHt}b(\vq)e^{-iHt}
\end{equation}
with $H$ as in \eqref{H_onetime}. Then one can show that 
\begin{equation}\label{phiaPsi}
\phi \bigl( x^{4M},y^{4N} \bigr) = \frac{(-1)^{M(M-1)/2}}{\sqrt{M!N!}}\scp{\emptyset}{a(x_1)\cdots a(x_M)b(y_1)\cdots b(y_N)|\Psi}
\end{equation}
on spacelike configurations, where $|\emptyset\rangle$ is the Fock vacuum. Equivalently, one can write using the field operators $\Phi_x = a + a^\dagger$, $\Phi_y = b + b^\dagger$,
\begin{equation}\label{phiPhiPsi}
\phi \bigl( x^{4M},y^{4N} \bigr) = \frac{(-1)^{M(M-1)/2}}{\sqrt{M!N!}} \scp{\emptyset}{\Phi_x(x_1)\cdots \Phi_x(x_M)\Phi_y(y_1)\cdots \Phi_y(y_N)|\Psi}
\end{equation}
on \emph{collision-free} spacelike configurations, i.e., those where none of the $x_j$'s and $y_k$'s are equal. 

In fact, we could take \eqref{phiPhiPsi} as the definition of $\phi$ on the collision-free spacelike configurations. Note that \eqref{phiPhiPsi} would define some multi-time function $\tilde\phi$ also for configurations with collisions, and even for non-spacelike configurations. In the absence of interaction, $\tilde\phi$ agrees with $\phi$, but in the presence of interaction the two differ at collision configurations, and $\phi$ is not defined at non-spacelike configurations; in that case, $\tilde\phi$ has the disadvantages that it does not necessarily satisfy any system of PDEs and that it is not related in a simple way to detection probabilities, as the curved Born rule \eqref{curvedBorn} holds only on spacelike configurations. (In addition, in the Heisenberg picture the Hilbert space $\Hilbert$ and the state vector $\Psi\in\Hilbert$ refer to the initial time $t=0$, and thus to a particular spacelike hypersurface. In contrast, $\phi$ and the PDEs \eqref{mt_emab} governing it are independent of any choice of hypersurface, so they provide a more fully covariant description.)

\subsection{Relation to the Tomonaga-Schwinger picture}\label{sec:TS}

The Tomonaga-Schwinger approach associates a wave function $\tilde\psi_\Sigma$ with every spacelike hypersurface $\Sigma$. We have pointed out in \eqref{phisigma} and \eqref{usigma} how multi-time equations define $\phi_\Sigma$, here
\begin{equation}
\phi_{\Sigma} \in \Hilbert_{\Sigma} = \bigoplus_{M=0}^\infty S_x L^2(\Sigma,\CCC^4)^{\otimes M} \otimes \bigoplus_{N=0}^\infty S_y L^2(\Sigma,\CCC^4)^{\otimes N},
\end{equation}
and unitaries $U_\Sigma^{\Sigma'}$. They are closely related to the Tomonaga-Schwinger approach, except that the latter is formulated in the interaction picture, which arises if we would like to represent $\phi_{\Sigma}$ by vectors $\tilde{\psi}_{\Sigma}$ in a fixed Hilbert space $\tilde{\Hilbert}$. Then, we need to identify each $\Hilbert_{\Sigma}$ with $\tilde{\Hilbert}$. This can be done via the free time evolution
\begin{equation}
F_{\Sigma\to\Sigma'} = \left( \bigoplus_{M=0}^\infty F_{\Sigma\to\Sigma'}^{(1,x)\, \otimes M} \right) \otimes \left( \bigoplus_{N=0}^\infty F_{\Sigma\to\Sigma'}^{(1,y)\, \otimes N} \right),
\end{equation}
where $F_{\Sigma\to\Sigma'}^{(1,x / y)}$ is the unitary operator obtained from solving the free Dirac equation with mass $m_{x / y}$. Then
\begin{equation}\label{tildepsi}
\tilde{\psi}_{\Sigma} = F_{\Sigma\to\Sigma_0} \phi_{\Sigma}
\end{equation}
is the wave function on $\Sigma$ in the interaction picture,
where $\Sigma_0$ is some fixed spacelike hypersurface connected to the choice of $\tilde{\Hilbert}$. The evolution of $\tilde\psi_\Sigma$ is given by the \emph{Tomonaga-Schwinger equation}
\begin{equation}\label{TS}
i \left(\tilde{\psi}_{\Sigma'} - \tilde{\psi}_{\Sigma}\right) = \left( \int_{\Sigma}^{\Sigma'} d^4x H_I(x) \right) \tilde{\psi}_{\Sigma},
\end{equation}
for infinitesimally neighboring spacelike hypersurfaces $\Sigma,\Sigma'$, where the integral is understood to be over the 4-dimensional volume enclosed between $\Sigma$ and $\Sigma'$, and where $H_I(x)$ is the Hamiltonian density in the interaction picture. For the model \eqref{Hams}, it is given by
\begin{equation}\label{emab_H_I}
H_I(x) = e^{i(H_x+H_y)x^0} a^\dagger(\vx) \Big( g^*\cdot b(\vx) + g\cdot b^\dagger(\vx) \Big) a(\vx) e^{-i(H_x+H_y)x^0}.
\end{equation}
The Tomonaga-Schwinger equation \eqref{TS} has a solution for every initial datum if and only if the consistency condition
\begin{equation}\label{consistency_TS}
\big[ H_I(x),H_I(y) \big] = 0 ~~\text{for all spacelike separated}~x,y
\end{equation}
holds. Furthermore, \eqref{TS} is Lorentz invariant if $H_I(x)$ is a Lorentz scalar. Note that \eqref{emab_H_I} is not a Lorentz scalar due to the choice of a fixed spinor $g$.

For the model \eqref{mt_emab}, one can show that the $\tilde{\psi}_{\Sigma}$ obtained from $\phi$ through \eqref{tildepsi} indeed solves the Tomonaga-Schwinger equation \eqref{TS} with interaction Hamiltonian density \eqref{emab_H_I}.
Conversely, any given solution $\tilde\psi_{\Sigma}$ to the Tomonaga-Schwinger equation \eqref{TS} with Hamiltonian density \eqref{emab_H_I} is connected to a solution $\phi$ of the multi-time equations \eqref{mt_emab}: (i)~we can switch from the interaction picture to $\Hilbert_\Sigma$ by considering $\phi_\Sigma := F_{\Sigma_0\to\Sigma} \tilde \psi_\Sigma$; (ii)~as mentioned before, Eq.~\eqref{objectivitycond} is the condition for the possibility of combining all the $\phi_\Sigma$ into a multi-time function $\phi$; (iii)~one can check \cite{pt:2013c} that \eqref{objectivitycond} is indeed satisfied; and (iv)~the multi-time evolution of $\phi$ agrees with \eqref{mt_emab} \cite{pt:2013c}. An analogous translation between Tomonaga-Schwinger equations for $\tilde\psi_\Sigma$ and multi-time equations for $\phi$ persists under rather weak assumptions on $\Hilbert_{\Sigma}$ and $H_I(x)$.

\subsection{Other models}

In \cite{pt:2013d}, we set up a multi-time model involving three particle species $x$, $y$ and $z$. In this model, $x$- and $y$-particles can annihilate each other and create a $z$-particle, and, conversely, a $z$-particle can decay into an $x$- and a $y$-particle. The interaction Hamiltonian is
\begin{equation}
H_{\mathrm{int}} = \int d^3\vx \Big( g a^{\dagger}(\vx)b^{\dagger}(\vx)c(\vx) + g^* a(\vx)b(\vx)c^{\dagger}(\vx) \Big),
\end{equation}
where $a,b,c$ are spinor-valued annihilation operators, and $g\in (\CCC^4)^{\otimes 3}$ (i.e., summation about spinor indices is implicitly understood). This model is inspired by electrons ($x$), positrons ($y$) and photons ($z$), but, as in the emission-absorption model, we take all three particles to be Dirac particles. The multi-time equations can be set up similarly as in \eqref{mt_emab} and \eqref{part_Ham}. It turns out this model is consistent on spacelike configurations if and only if the fermion number is conserved, i.e., either none or two of the $x,y,z$ particles are fermions. It can be shown that this model is related to quantum fields in the Heisenberg picture and the Tomonaga-Schwinger picture in the same way as described above.

We conjecture that multi-time equations can be set up consistently for many kinds of QFTs under some reasonable conditions, and that the equivalence to quantum fields in the Heisenberg picture and the Tomonaga-Schwinger approach still holds.

\section{Conclusions}

In this paper, we have given an overview of the theory of multi-time wave functions and its recent developments. It was elucidated that the consistency of the multi-time evolution is a restrictive condition that excludes the most common mechanism of interaction in non-relativistic quantum mechanics, i.e., potentials. We have described two alternative ways of constructing relativistic interactions in the multi-time picture: zero-range (or $\delta$-potential) interactions and interactions via particle creation and annhilation in QFTs. We have also described the relations of the multi-time wave function to detection probabilities along spacelike hypersurfaces, to the field operators in the Heisenberg picture, and to the Tomonaga-Schwinger approach.

One striking trait of the multi-time approach lies in its parallels to non-relativistic quantum mechanics: in that the quantum state is represented by a \emph{wave function}, that its time evolution is governed by PDEs, and that its modulus squared yields detection probabilities. And the results reported here suggest  that it may be possible to formulate also more serious relativistic QFTs in terms of multi-time wave functions, which sets a goal for future research.

\section*{Acknowledgments}

\begin{minipage}{15mm}
\includegraphics[width=13mm]{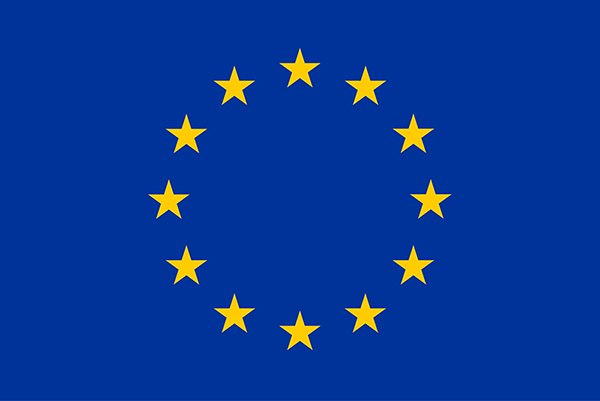}
\end{minipage}
\begin{minipage}{143mm}
This project has received funding from the European Union's Framework for Research and Innovation Horizon 2020 (2014--2020) under the Marie Sk{\l}odowska-
\end{minipage}\\
Curie Grant Agreement No.~705295. 
S.P.\ gratefully acknowledges support from the German Academic Exchange Service (DAAD).

\end{document}